# An Accurate Method for Measuring Activation Energy

**Xiang-Bai Chen\*, Jesse Huso, John L. Morrison, and Leah Bergman**

**Department of Physics, University of Idaho, Moscow, ID 83844-0903**

## Abstract

In this letter, we present an accurate method for the measurement of activation energy. This method combined the excitation power dependent photoluminescence and temperature dependent photoluminescence together to obtain activation energy. We found with increasing temperature, there is a step transition from one emission mechanism to another. This step transition gives us an accurate measurement of activation energy. Using this new method we found the activation energy of the free exciton *A* in a GaN thin film is 24±1 meV. Our result also gives a reasonable explanation of the debate of the origin of the light emission in GaN at room temperature.

\* Electronic address: chen7779@uidaho.edu

Activation energy is a very important parameter determining the mechanism of optical transitions, which predominantly determine the performance of optoelectronic devices such as light emitting diodes, laser diodes, and detectors. For example, in the room temperature optoelectronic device applications using the wide bandgap semiconductor GaN, it is very important to know if the free exciton *A* transition is still dominant at room temperature. Thus it is desirable to obtain an accurate measurement of the activation energy for excitons, since the activation energy of excitons gives a good measure of their thermal stability. A common method for the measurement of activation energy is through the study of temperature dependent photoluminescence, which using the photoluminescence intensity as a function of temperature to determine activation energy [1-7]. In this letter, we presented a new method combining the excitation power dependent photoluminescence and temperature dependent photoluminescence to measure activation energy. First, we applied the excitation power dependent photoluminescence to determine the mechanism of the optical transition at a certain temperature, then we applied the temperature dependent photoluminescence to obtain the information of optical transitions at all temperatures. At the temperature whose thermal energy is comparable to the activation energy, we should observe a transition between the optical emission mechanisms, and this thermal energy gives us the information of activation energy. In the experiment, we did observe a step transition, which gives an accurate measurement of activation energy.

Our experiments utilized a JY-Horiba micro-photoluminescence system consisting of a high-resolution T-64000 triple monochromator and a UV microscope capable of focusing a spot size of ~ 1 μm in diameter. Additionally, the system has a CCD camera and a monitor, enabling the viewing of the laser spot as well as the area sampled. A He-Cd laser with a wavelength of 325 nm (3.8 eV) was employed as the excitation source. The low and high temperature photoluminescence measurements were carried out utilizing an Instec micro-cell, in the temperature range of 183 K to 403 K. The GaN thin film used in our experiment was grown via MOCVD on 6H-SiC (0001) substrates, the thickness of the films is ~ 1μm.

The excitation power dependent photoluminescence is widely used for determining the origin of light emission in semiconductors. It has been established that the luminescence intensity, $I$, can be expressed as [8-10]:

$$I = \eta I_0^\alpha \quad (1)$$

In this relation $I_0$ is the power of the exciting laser radiation, $\eta$ is the emission efficiency, and the exponent α represents the radiative recombination mechanism. For free exciton emission the value of α has been reported in the range 1-1.3 [8, 11-17]; for bandgap emission, i.e. free-electron-hole recombination, α = 2 [8, 11, 17]. Figure 1 presents the photoluminescence intensity as a function of the excitation laser power for a GaN thin film at the temperature of 183 K. The data are well described by equation (1) with a line of slope 1.18 on the logarithmic plot, indicating the exponent α = 1.18, implying free exciton *A* transition in the GaN thin film at that temperature.

Above we used the excitation power dependent photoluminescence study obtained the information of the origin of the light emission for a certain temperature. In order to gain further information, we combined the temperature dependent photoluminescence study with the excitation power dependent photoluminescence study. In Figure 2, we present the exponent α as a function of temperature from 183 K to 403 K. As can be seen in the figure, when temperature increased from 183 K to 263 K, the exponent α ~ 1.2; when temperature increased from 263 K to 283 K, there is a step jump of the exponent α from ~ 1.2 to ~ 1.6; when temperature increased from 283 K to 403 K, the exponent α ~ 1.6-1.8. Our results indicate that in the temperature range 183-263 K, the light emission is from free exciton *A*; while in the temperature range 283-403 K, the emission is mainly due to bandgap recombination; and at a temperature between 263-283 K there is a transition from the free exciton emission to bandgap recombination.

The most interesting and useful observation from Figure 2 is that there is a step jump of the value of exponent α from 263 K to 283 K. This phenomenon provides a new method to calculate the activation energy of exciton more accurately. We conclude that the activation energy of the exciton can be calculated by:

$$E_A = k_B T \pm k_B \Delta T \qquad (2)$$

where *T* is the transition temperature, which is a temperature between 263-283 K, and $k_B \Delta T$ is the experimental error, the $\Delta T$ can be taken as the half width of the step. Due to the step jump of the value of exponent α as a function of the temperature, this method provides a very accurate measurement for the activation energy. Normally, the activation energy is obtained through the temperature dependent photoluminescence intensity, which can be described by [1-6]:

$$\frac{I(T)}{I_0} = \frac{1}{1 + C \exp(-E_A / k_B T)} \qquad (3)$$

where $E_A$ is the activation energy, *I(T)* is the photoluminescence intensity at temperature *T*, $I_0$ is the photoluminescence intensity at absolute zero temperature, and *C* is a constant. When equation (3) is used to calculate the activation energy, there are two restrictions: first $C \exp(-E_A / k_B T) \gg 1$, and second the photoluminescence intensities at different temperatures has to be measured consistently. Due to the thermal expansion, the second requirement may introduce a significant error, especially for micro-photoluminescence experiments. With the new method, though neither the restriction is required, also there is a step jump of the value of the exponent α. Therefore we conclude the new method gives a more accurate measurement of the activation energy.

Using equation (2), we found the activation energy of the free exciton *A* in the GaN thin film is 24±1 meV. In the literature, values in the range 18-30 meV of the activation energy for the free exciton *A* have been reported [1-4, 18-19]. Due to the wide range of the reported values of the activation energy, there was a debate on the origin of the light emission from GaN at room temperature [20]. Some authors observe the free exciton *A* emission [21-22], while other authors observe the band to band transitions [23-24]. Our new method provides an accurate measurement of the activation energy, and also gives a reasonable explanation of the debate. Since the activation energy of free exciton *A* is very close to the thermal energy at room temperature, and due to the step function in the transition of the light emission mechanisms, when considered with slight differences between samples, the origins of light emissions at room temperature can be different. The observation of different transitions in different samples has been reported [25]. Thus we expect that the debate of the origin of the light emission in GaN at room temperature is mainly due to the sample difference in different groups and the not accurate measurement of activation energy.

In conclusion, we presented an accurate new method for measuring the activation energy. This method combined the excitation power dependent photoluminescence and the temperature dependent photoluminescence to measure activation energy. Using our new method, we found the activation energy of the free exciton *A* in a GaN thin film is 24±1 meV. Also, our results gave a reasonable explanation for the debate over the origin of the light emission in GaN at the room temperature.

**Acknowledgments**

The authors gratefully acknowledge NSF CAREER DMR-0238845 and DOE-DE-FG02-04ER46142, as well as the American Chemical Society PRF 40749-AC10.

**Figure Captions:**

Figure 1. Photoluminescence intensity on logarithmic scale as a function excitation laser power at the temperature 183 K.

Figure 2. The exponent α as a function of temperature.

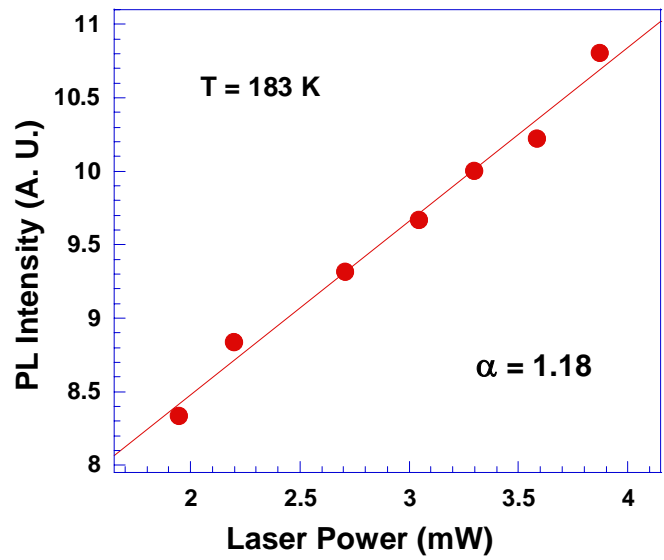

**Figure 1**

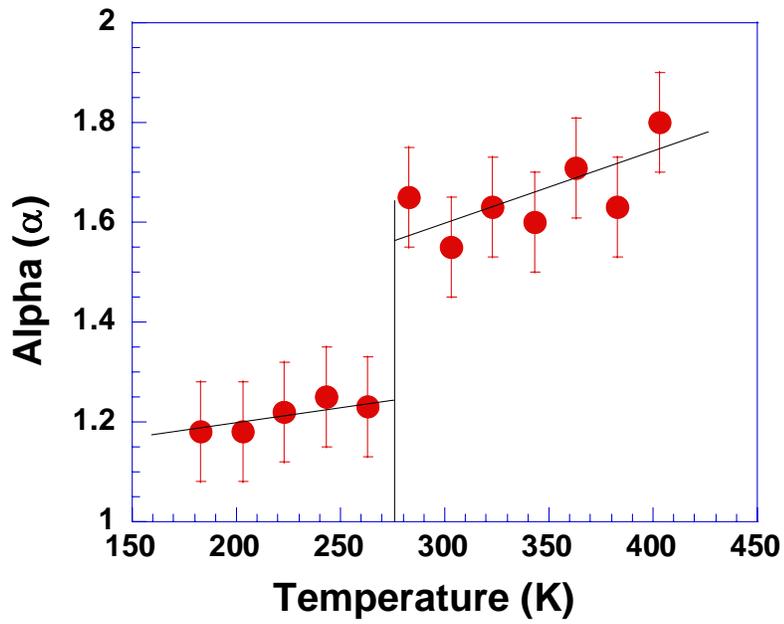

**Figure 2**